\documentclass{aastex}

\usepackage{spr-astr-addons} 
\begin{document}
\title{Implications of an Anti-glitch in AXP/SGR}
\shorttitle{Anti-glitch}
\shortauthors{Katz}
\author{J. I. Katz\altaffilmark{}}
\altaffiltext{1}{Department of Physics and McDonnell Center for the Space
Sciences \\ Washington University, St. Louis, Mo. 63130}
\begin{abstract}
The recently observed anti-glitch of AXP 1E 2259+586 may be explained as the
consequence of sudden accretion of retrograde matter.  AXP/SGR are explained
as single neutron stars accompanied by fallback matter from their natal
supernov\ae, including rocky or metallic planetesimals.  The upper bounds on
a glitch or anti-glitch in the giant outburst of SGR 1806-20 pose a problem
for any model.
\end{abstract}
\maketitle
\section{Introduction}
Dissipation of the magnetostatic energy of a neutron star with magnetic
field greater than those of radio pulsars was suggested \citep{K82} to
explain the giant outburst of March 5, 1979 from a source (now known as SGR
0526-66) in a young supernova remnant in the LMC.  It was further developed,
and the name ``magnetar'' coined, by \cite{DT92,TD95} and subsequent work;
see \cite{M08} for a review.  The irregular and unpredictable behavior of
Solar and stellar flares, also powered by magnetic dissipation, is then a
model of the behavior of soft gamma repeaters (SGR).  The amount of energy
released in their most energetic outbursts requires \citep{K82} magnetic
fields far in excess of $10^{12}$--$10^{13}$ gauss, an inference apparently
confirmed by the subsequent measurement of the periods (5--12 s) and 
spin-down rates of their steady emission, during which they are called
anomalous X-ray pulsars (AXP).  It is generally believed that AXP and SGR
are different aspects of the same objects \cite{M11}.

The magnetar model attributes the emission of radiation, steady but 
periodically modulated (in AXP) or episodic (in SGR), to magnetic
dissipation.  Its fundamental assumption is that an isolated neutron star
loses angular momentum at the rate indicated by classic pulsar theory
\citep{GJ69}, describing the interaction of the neutron star with its 
environment as the same as that of radio pulsars in vacuum.  This assumption
permits the magnetic dipole moment to be inferred from the spin-down rate,
and is the basis of the inference that AXP/SGR have much larger magnetic
fields and magnetostatic energy than classic radio pulsars (hence the label
``magnetar''), sufficient to power both the steady emission of AXP and SGR
outbursts.  A natural prediction that follows from this fundamental
assumption is that the periods of AXP, like those of radio pulsars, will
steadily increase as they lose angular momentum.  Intermittent coupling to
a superfluid component can explain sudden period decreases (``glitches'') in
AXP as it does in radio pulsars.

The picture was complicated by the discovery that some AXP have magnetic
dipole moments (as inferred from their spin-down rates) within the range of
those of radio pulsars.  Something other than magnetic dipole moment must
distinguish AXP (and by implication SGR) from those neutron stars that, if
spinning fast enough, would be radio pulsars.  In a different class of
models \citep{PTH95} this difference is attributed to the presence in
AXP/SGR of a fallback disc of matter from the event that produced the
neutron star.
\section{The Anti-glitch}
\cite{A13} discovered an ``anti-glitch'' in the AXP 1E 2259+586.  In
contrast to the glitches of radio pulsars and other AXP, in the anti-glitch
the rotation rate suddenly decreased (spin-down).  In all previously known
glitches the rotation rate increased (spin-up), which is explained as a
sudden coupling of a more rapidly rotating superfluid component to the
observed solid crust that steadily loses angular momentum to external
radiation torque.

In any model based on an isolated rotating object, such as a classical
single radio pulsar or the hypothesized magnetar, in which rotation slows
steadily, closer coupling to a superfluid component (or to any weakly
coupled rotating component) can only increase the rotation rate, because a
weakly coupled component must be rotating faster than the crust on which
slowing torques are exerted.  This is the generally accepted explanation of
spin-up glitches.  In contrast, an anti-glitch would require coupling to a
component rotating more slowly than the crust was rotating before the
anti-glitch, and hence more slowly than the crust has {\em ever} rotated
since the object's birth.

Equally remarkable, after all previously known glitches the spin-down rate
increased, the rotation rate partly relaxing to its pre-glitch spin-down
trajectory.  This is explained as a reduction in the crust-core coupling
following its sudden increase (and presumed reduction of the difference in
their rotation rates) in the glitch; there may be an equilibrium frequency
lag, with deviations from it (like those following a glitch) gradually
relaxing.  In contrast, the anti-glitch of 1E 2259+586 was followed by an
increase in the steady spin-down rate, {\em increasing} the discrepancy
between its rotation rate and an extrapolation of its pre-anti-glitch
slowing.  This is inconsistent with a weakly coupled reservoir of angular
momentum whose torque on the crust is a monotonic function of the difference
in their rotation rates.
\section{Magnetar Models}
Despite their popularity, there are unresolved problems with magnetar
models.  Magnetar models are based on the inference of extraordinarily large
magnetic fields from the classic pulsar relation \citep{G68,GJ69} between
magnetic dipole moment and spindown.  In this model a magnetar's spin-down
rate should be very stable because it is proportional to the square of its
magnetic dipole moment.  The X-ray pulse profile should also be stable
because it is determined by the structure of the magnetic field, unlike the
pulse profiles of radio pulsars that are determined by complex plasma
processes, and those of accreting X-ray pulsars that are determined by the
magnetohydrodynamics of plasma penetration into the magnetosphere.  These
predictions of the magnetar model must be reconciled with the observation of
${\cal O}(1)$ changes in the spindown rates and pulse profiles of SGR
1900+14 \citep{MRL99,W99,P01}, AXP 4U0142+61 \citep{DKG07,DKG08}, and AXP
1E 1048.1-5937 \citep{DKG09}, among others.

The stable spin-down rates of radio pulsars show that neutron star magnetic
dipole moments, and any other properties that affect spin-down, do not
change on short time scales.  If neutron star magnetic fields could reorder
themselves freely, energy minimization would long ago have reduced the
dipole moments to small values.  Even an SGR outburst should only change the
magnetic configuration and dipole moment by ${\cal O}(1\%)$.  This follows
theoretically from comparison of the outburst energy to the magnetic energy
inferred from the spin-down rate.  Empirically, it follows from the fact
that typical intervals between outbursts appear to be several decades or a
century, but SGR are generally found in SNR several thousand years old,
implying that they undergo ${\cal O}(100)$ major outbursts in their lives.

A number of modifications \citep{TLK02,TXSQ13} of the magnetar model have
been suggested in order to resolve these issues.  For example, \cite{L13}
and \cite{To13} have suggested that AXP timing irregularities and the newly
discovered anti-glitch may be reconciled with magnetar models.  They argue
that opening of magnetic field lines to infinity, in analogy to Solar
Coronal Mass Ejections or the Solar wind, might increase the spindown torque
by as much as nine orders of magnitude.  Such an event, observed with the
low cadence of X-ray monitoring of AXP, could appear as an anti-glitch.  The
present understanding of magnetospheres is insufficient to decide whether
this is a satisfactory explanation.  The fact that such events were
not contemplated in magnetar theory prior to their discovery, and are not
observed for radio pulsars that are the paradigms of magnetars, suggests
that other explanations should be considered.
\section{Accretional Models}
Accretion, satisfactorily explaining X-ray emission from neutron
stars in binary systems, has been suggested as an alternative explanation
of AXP \citep{PTH95} and even of the outbursts of SGR \citep{KTU94,K96}.
In these models the outbursts are produced by an optically thick photon-pair
plasma \citep{K96}, just as they may be in magnetar models \citep{K82}.
Such a plasma has a characteristic photospheric temperature and brightness
determined by the requirement that an equilibrium pair plasma be optically
thick to Compton energy equilibration.

The chief observational objection to accretional models is the extreme
faintness (in most cases, undetectability) of visible or infrared
counterparts to AXP \citep{M08}, such as might be expected to be produced by
the cooler regions of an accretion disc.  However,
\cite{CHN00,A01,M08,E09,M10,T13,ACE13} have considered models based on
supernova fall-back, consistent with the association of AXP and SGR with
young SNR.  Such fall-back would be composed of matter from the deep
interior of the pre-supernova star, which would have undergone much nuclear
processing, perhaps to iron.  It would (unlike hydrogen-dominated 
mass-transfer discs) readily condense to discrete rocky or iron planets or
planetesimals.  The visible and infrared luminosity of such matter would be
very small because its total surface area would be much smaller than that
of a continuous fluid accretion disc.

\cite{KTU94} suggested that accretion of objects analogous to the planets
orbiting at least one radio pulsar \citep{WF92}, rather than of a continuous
disc, might explain both X-ray/$\gamma$-ray outbursts, whose temporal
behavior is reminiscent of meteor showers, and faintness at visible and
infrared wavelengths.  Evidence for accretion of discrete solid
planetesimals by white dwarves \citep{DWS12} and for the presence of an
``asteroid belt'' around a radio pulsar \citep{S13} have been reported.  In
this model accretion occurs when collision or gravitational interaction
produces fragments with almost zero angular momentum that fall onto the
compact star.  The residual angular momentum may be either prograde or
retrograde, allowing for anti-glitches.  The dominant sense of angular
momentum in fallback matter is likely to be that of the core of the
supernova that produced it, the same as the sense of rotation of the remnant
neutron star.  Most glitches resulting from accretion of condensed
(asteroidal) objects are therefore likely to be spin-up glitches, although
subsequent relaxation may differ from that of the internally-driven glitches
of isolated radio pulsars.  Despite this, discrete gravitational or
collisional impulses \citep{KTU94} can place some objects on orbits with
small retrograde angular momentum, leading to anti-glitches when accreted by
the neutron star.

Analogy to binary neutron star X-ray sources implies that in accretional
models the torques exerted on the neutron star by surrounding matter may
produce either spin-up or spin-down, may vary irregularly, and may be much
larger than the electromagnetic torque on an isolated neutron star.  In
an accretional model the magnetic field cannot be inferred from the spindown
rate with the usual pulsar relation, and it may be smaller than implied by
the magnetar model.
\section{Application to AXP 1E 2259+586}
\label{2259+586}
In an accretional model the anti-glitch of 1E 2259+586 is attributed to the
sudden accretion of matter with the opposite sense of angular momentum as
the neutron star's rotation, or to a transient increase in the ``propeller''
interaction of the neutron star's magnetosphere with surrounding matter.
Sudden accretion may be attributed to the infall of a solid body.   Some of
its matter may be rapidly accreted, but some may remain in a nearby
accretion disc from which it is more gradually accreted.  This may be the
the origin of the steadier spin-down torque as well as the increase in
luminosity following the anti-glitch.  The variability of accretion torques
on a range of time scales, known from the spin histories of binary neutron
star X-ray sources, may explain the changing spin-down rates of AXP.  Some
of the disc matter may be expelled, carrying away angular momentum, but some
of it may be accreted, powering the steady X-ray emission.  In such a model
the dipole field cannot be inferred from the spin-down rate. 
 
The anti-glitch of AXP 1E 2259+586 is readily explicable in accretional
models.  The specific angular momentum of accreted matter may be estimated
from the properties of the source in its steady state:
\begin{equation}
\label{elleq}
\ell = {I_{NS} \vert {\dot \omega} \vert \Omega_{NS} \over L} \approx 6
\times 10^{17} {\rm cm^2/s},
\end{equation}
where the luminosity $L$ is obtained from the observed steady soft X-ray
flux \citep{A13} and the distance \citep{KF12}, $\dot \omega$ is the steady
spin-down rate \citep{A13}, the moment of inertia $I_{NS} = 1.5 \times
10^{45}$ g-cm$^2$ and the gravitational binding energy $\Omega_{NS} \approx
10^{20}$ erg/g \citep{RP94}. 

For a discrete accretional event, such as the anti-glitch, the accretional
energy 
\begin{equation}
E = M \Omega_{NS} = {I \vert \Delta \omega \vert \Omega_{NS} \over \ell},
\label{Eeq}
\end{equation}
where $M$ is the accreted mass and $I$ is the moment of inertia to which the
accreted mass is coupled during the anti-glitch; in general $I \le I_{NS}$
because of the weak coupling of the solid crust to the superfluid component.
\cite{A13}, in their Model 1, find the anti-glitch magnitude to have been
$\Delta \omega \approx - 3 \times 10^{-7}$/s.

Because neutron stars have a weakly coupled superfluid component, the value
of $I$ applicable to a transient impulsive torque, such as that resulting
from the accretion of a discrete body, may be only that of its solid
component $I_s \approx 2 \times 10^{43}$ g-cm$^2$ \citep{RP94,HG13}.  Taking
this value implies $E \approx 10^{39}$ erg and $M \approx 10^{19}$ g for the
anti-glitch of AXP 1E~2259+586.  Gradual coupling of the solid to the
superfluid components may account for complex post-glitch spin behavior, as
is observed following the glitches of radio pulsars.  Similar effects may be
produced if some of the discrete body is not accreted immediately, but more
gradually from an accretion disc.

It is not possible to determine over what duration $\tau$, within the
observationally constrained interval of about $10^6$ s, this energy was
released.  At a distance of 3.2 Kpc (\cite{KF12}; this has been
controversial, and \cite{DK06} and \cite{T10} suggested larger values) the
fluence corresponding to $E$ is $\approx 10^{-6}$ erg/cm$^2$ and the flux
$\approx 10^{-12}(10^6\,{\rm s} /\tau)$ erg/cm$^2$-s.  This fluence is about
20 times that measured for a 36 ms hard X-ray burst consistent with the time
of the anti-glitch by the Fermi GRB Monitor \citep{A13}, but the hard X-ray
fluence may require a large bolometric correction.

The observed \citep{A13} decaying increment to the 2--10 keV X-ray flux
following the anti-glitch implies an incremental radiated energy $\approx
5 \times 10^{40}$ ergs, more than suggested by $I = I_s$ but less than by
the total moment $I_{NS} \approx 1.5 \times 10^{45}$ g-cm$^2$ \citep{RP94}.  
This is consistent with partial relaxation of the angular momentum impulse
to the superfluid component, with some residual increased differential
rotation.  The more rapid spindown during this period of increased flux is
also consistent with continuing accretion of retrograde matter, and the
decay of the flux increment is consistent with the return of the spindown
rate to approximately its pre-anti-glitch value in the latter half of 2012.
\section{Discussion}
Both accretional and magnetar models predict a correlation of the magnitude
of the spin-down rate with luminosity, both during and outside glitches (or
anti-glitches), as observed for 1E~2259+586 \citep{A13}.  In general, the 
spin-down rates are larger in accretional models because accreted matter is
only semi-relativistic and has a greater ratio of momentum to energy than
radiation or relativistic particles.  Only accretional models offer a
natural explanation of anti-glitches.

These estimates may be applied to the giant outbursts of SGR.  SGR 1900+14
underwent an outburst August 27, 1998 during which its spin rate may have
decreased by $\Delta \omega \approx - 1 \times 10^{-4}$/s \citep{W01}.  This
may be considered to have been a giant anti-glitch, and, if so, its magnitude
was consistent, allowing for uncertainty in $\ell$ and $I_s$, with the
emission of $10^{44}$ ergs \citep{M08} powered by accretion.  SGR
1806-20 underwent an outburst 100 times more energetic on December 27, 2004,
but for it $\vert \Delta \omega \vert \lesssim 4 \times 10^{-6}$/s
\citep{W07}, inconsistent with both models, even if $I$ is the full stellar
moment of inertia $I_{NS}$.   In any model, or phenomenologically, it is
difficult to reconcile the disparate rotation changes during these two giant
outbursts unless the result for SGR 1900+14 is considered only an upper
limit.

In a magnetar model $\ell \approx c r \approx 3 \times 10^{16} (r/r_{NS})$ 
cm$^2$/s, where $r$ is the radius of emission, unless matter is expelled and
radiation emitted symmetrically so that the net angular momentum lost is 
small.  For the outburst of SGR 1806-20, $E \gtrsim 1.6 \times 10^{46}$ ergs
\citep{M08}, implying $\Delta \omega \gtrsim 3 \times 10^{-4} (r/r_{NS})$/s,
even if only radiation and relativistic matter are emitted (corresponding to
$\Omega_{NS} \approx c^2$ in Eq.~\ref{Eeq}), at least 100 times greater than
the observational upper limit.  The luminosity and spectrum \citep{M08}
imply $r \gtrsim 10^2 r_{NS}$ and $\ell \gtrsim 3 \times 10^{18}$ cm$^2$/s,
even if emitting as a black body, tightening the requirement on the symmetry
of emission to roughly one part in $10^4$.

An analogous problem arises in almost any model of SGR 1806-20, including
accretional models, because emitted radiation carries momentum and angular
momentum.  Very generally, 
\begin{equation}
\vert \Delta \omega \vert \gtrsim {E r \over I c} \approx 3 \times 10^{-4}
{E_{46} r_6 \over I_{45}} {\rm s}^{-1},
\end{equation}
where $E_{46} \equiv E/(10^{46}$ ergs), $I_{45} \equiv I/(10^{45}$ g-cm$^2$)
and $r_6 \equiv r/(10^6$ cm). 
Only if radiation (or a particle wind) is emitted uniformly across the
neutron star surface, or if the emitting object, such as a white dwarf
\citep{MRR12}, has a much larger value of $I$, can the problem be avoided.  



The observation of the anti-glitch in AXP 1E 2259+586 supports the
hypothesis that neutron stars in environments intermediate between the
interstellar vacuum of radio pulsars and the dense plasma of mass transfer
binary X-ray sources become AXP/SGR, as suggested by \cite{A01}.  They may
be surrounded by, and accrete, sufficient residual matter, condensed as well
as gaseous, from their natal SN (consistent with their location in young
SNR) to slow their rotation and increase their spin-down rates to the
observed values, and to power their steady emission.  Without stellar
companions they do not have the luminous plasma accretion discs and rapid
accretional spin-up of neutron stars in binary systems, and ``propeller''
spindown may occur instead.  This intermediate density environment may be
the critical factor distinguishing them from those neutron stars that, when
spinning more rapidly, are radio pulsars.
\appendix
\section{Anti-glitch in PSR B0540$-$69 (added May 5, 2024)} 
\cite{Tuo24} reported an anti-glitch in PSR B0540$-$69.

Another possible cause of an anti-glitch is the accretion of a planetesimal
from an orbit with angular momentum whose dot-product with tne PSR
angular momentum is negative.  This would imply no change in the spin-down
rate after the glitch (unless the magnetic configuration were altered,
which might be unlikely).  That is consistent with the results of
\cite{Tuo24} showing $\Delta{\dot\nu}$ of $-1.2\sigma$, not significantly
different from zero.

Our paper \cite{KTU94} suggests a model of SGR in which they are
caused by accretion of planetesimals onto neutron stars (Ali Alpar published
the basic idea several years earlier).  In such a model the accretion would
produce an X-ray flare.  The observed change in spin rate $\Delta\nu =
1 \times 10^{-7}$ Hz implies an accreted angular momentum of about $10^{38}$
g-cm$^2$/s.  Accreted matter directly impacting the surface carries a specific
angular momentum of about $1.5 \times 10^{16}$ cm$^2$/s, implying accretion of
about $10^{22}$ g, and gravitational energy release of about $10^{42}$ ergs.

However, it is more likely that a planetesimal would break up at a tidal
radius of about $10^{11}$ cm, and most such events would involve an orbit
with peri-astron about this value.  This would increase the specific angular
momentum of accreted matter to about $5 \times 10^{18}$ cm$^2$/s and reduce
the planetesimal's mass to about $2 \times 10^{19}$ g and the gravitational
energy release to about $2 \times 10^{39}$ ergs.  Because such an event would
create an accretion disc with Kepler time scale about 2000 s, and viscous
dissipation time scale one or more orders of magnitude longer, the emitted
power would plausibly be in the range $10^{34}$--$10^{36}$ erg/s, undetectable
at the range of the LMC.

A rocky planetesimal of $2 \times 10^{19}$ g would have a radius of about
10 km.  One PSR is known to be accompanied by planets, and it is plausible
that many other PSR are also accompanied by rocky debris (as are many WD).
Events like that which you have observed are a consequence (a prediction?)
of the SGR model of \cite{KTU94}.  While there are many ways to produce a
classical glitch with superfluid coupling, anti-glitches are harder to make.

\end{document}